\documentclass[aps,pra,a4paper,twocolumn]{revtex4-1}
\usepackage[utf8]{inputenc}
\usepackage[T1]{fontenc}
\usepackage{amsmath,amsthm,amssymb}
\usepackage{bbm}

\usepackage{graphicx}
\usepackage{tikz}
\usepackage{braket}
\usetikzlibrary{decorations.pathmorphing}
\usetikzlibrary{arrows}
\usetikzlibrary{matrix}

\usepackage{color}
\usepackage[normalem]{ulem}
\definecolor{gray}{rgb}{.6,.6,.6}
\definecolor{darkyellow}{rgb}{.6,.5,0}
\definecolor{darkgreen}{rgb}{0,.6,0}
\definecolor{darkblue}{rgb}{0,0,.6}

\usepackage{hyperref}
\usepackage{verbatim}

\begin{document}

\title{Quantum parameter estimation with a neural network}

\author{ Eliska Greplova $^{1}$, Christian Kraglund Andersen $^{2}$, Klaus Mølmer $^{1}$}
\affiliation{$^{1}$ Department of Physics and Astronomy, Aarhus University, Aarhus, Denmark}

\affiliation{$^{2}$ Department of Physics, ETH Zurich, Switzerland}

\date{\today}

\begin{abstract}
We propose to use neural networks to estimate the rates of coherent and incoherent processes in quantum systems from continuous measurement records. In particular, we adapt an image recognition algorithm to recognize the patterns in experimental signals and link them to physical quantities. We demonstrate that the parameter estimation works unabatedly in the presence of detector imperfections which complicate or rule out Bayesian filter analyses.
\end{abstract}

\maketitle

The development of quantum technologies with improved performance for a wide number of tasks rely on our ability to manipulate and control individual quantum systems which, in turn, assumes precise knowledge of system parameters such as transition frequencies, coupling strengths and dissipation rates. Operation of quantum communication and computation devices \cite{nielsen2010quantum} may require frequent verification and calibration measurements, while the whole purpose of quantum sensing \cite{RevModPhys.89.035002} and metrology \cite{PhysRevLett.96.010401} is to extract such parameters from measurement records.

For repeated measurements on a quantum system, the probability, or likelihood, for each measurement outcome is governed by the system density matrix or state vector which in turn depends on the parameters governing its dynamics, and sampling of $N$ independent measurement results leads to an estimation error that decreases as $1/\sqrt{N}$. For continuous measurements on a single quantum system, each random signal output is accompanied by measurement back action, which has consequences for future measurement outcomes. For many dissipative quantum systems, measurements are only correlated over finite time, and the estimation error decreases as $1/\sqrt{T}$ as function of the total duration, $T$, of the measurement~\cite{RevModPhys.68.1}. The optimal estimation of unknown parameters $\{\theta\}$ from a measurement signal is governed by the Cram\'er-Rao bound~\cite{cramer-rao} and the Fisher information~\cite{Fisher} which, in turn, involve the likelihood $L(D|\{\theta\})$ of the signal  $D$ conditioned on the values of the parameters in question~\cite{PhysRevA.87.032115}. If this likelihood function is known, Bayes' rule,
$L(\{\theta\}|D) \propto L(D|\{\theta\})\cdot P_{prior}(\{\theta\})$ saturates the Cram\'er-Rao bound and provides the optimal maximum likelihood estimate for $\{\theta\}$. But, it is in general very complicated to evaluate $L(D|\{\theta\})$, and hence to obtain an optimal parameter estimate when the signal is in the form of a continuous time dependent record $D=I(t)$.

\begin{figure}
\centering
\includegraphics[scale=0.26]{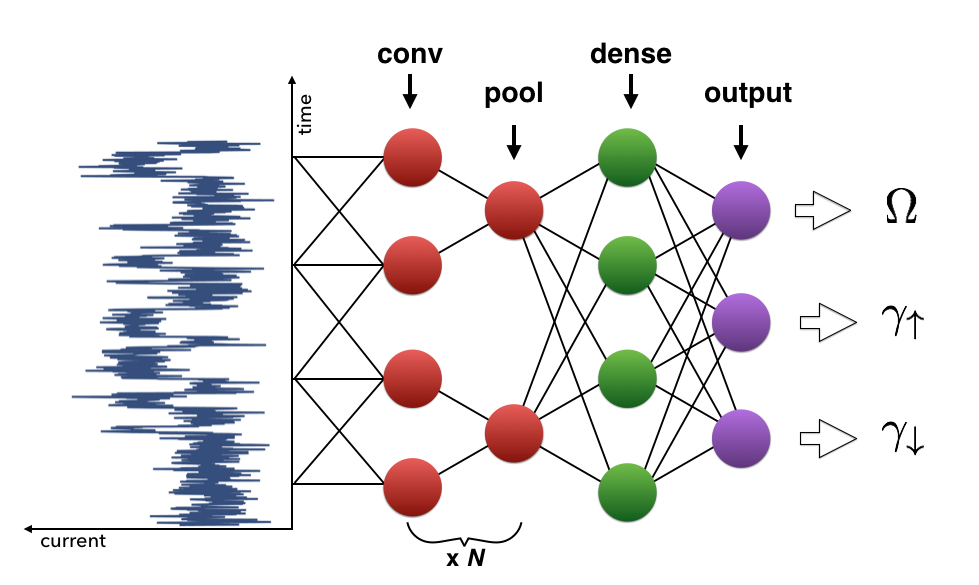}
\caption{The neural network used for parameter estimation based on analyses of experimental signals. From left to right, the figure shows how the experimental data are stored in an input layer of nodes that is connected to a convolutional layer which applies one or several series of filters followed by a pooling layer and a densely connected layer that identifies finer features in the data, and is then connected to the output. The output layer yields the probabilities and most likely candidate values of the parameters governing the production of each signal. The function of the neurons, shown as circles in the figure, is to forward output values that are simple functions of the inputs, and the purpose of the machine learning is to properly adjust these functions for optimal recognition on a large training set, see Supp. Mat.}\label{fig:NN}
\end{figure}

In this Letter, we propose to use an artificial neural network to analyze detector signals and identify the most likely set of parameters based on machine learning from independently generated training signals. Machine learning has recently found numerous applications in quantum physics~\cite{CarleoTroyer, PhysRevA.96.042113,PhysRevX.7.021021,PhysRevLett.119.030501} and it has been used for parameter estimation by projective measurements on quantum systems~\cite{QINFER}. Our work differs from these previous efforts by applying machine learning to the analysis of the weaker and more noisy measurements associated with continuous monitoring of an experimental system. Our measurement signal thus consist of a time series of data which we regard as a one-dimensional image in order to directly adapt machine learning algorithms developed for image recognition. Specifically, we employ convolutional neural networks, which are known to be highly effective for classification of images (see general introduction to machine learning and neural networks in Supplementary Material, ~\cite{goodfellow2016,Nielsen2015,florian,fei-fei-li}). The neural network is exposed to signals obtained for a finite number of parameter sets and the network is not provided with any information about how these parameters are associated with the dynamics of the system and the ensuing signals.


We have recently~\cite{Greplova2017} simulated sensing of the charge on a quantum dot by the noisy current probed by a quantum point contact (QPC). Under a Markovian assumption for the tunneling dynamics and the QPC current signal, the system was simulated and analyzed by a stochastic master equation~\cite{Molmer, WiMi2010}. By propagating quantum trajectories associated with each candidate set of parameters, we are able to continuously update the likelihood for each set in a Bayesian manner~\cite{PhysRevA.87.032115, PhysRevA.64.032111,PhysRevLett.108.170502,PhysRevA.64.042105,PhysRevA.69.032109}. Such Bayesian analysis becomes cumbersome for large parameter spaces and as shown in~\cite{Greplova2017}, it is also possible to combine Bayesian estimation of some parameters with a more straightforward frequency analysis of the others. Note, however, that these analyses make very specific assumptions about the dynamics and formation of the signal, i.e, they assume the validity of a stochastic master equation and perfect knowledge of all other parameters than the ones, that we wish to estimate. Inefficient detection and more complicated effects such as dead time and finite band width effects can be incorporated in the stochastic master equation and Bayesian theory, but only with considerable  effort, and only if their statistical properties are fully characterized~\cite{WW2003, WW2003-2, WWM2002}. There is, therefore, a need for methods that can effectively distinguish different physical cases without relying on assumptions on physical parameters that cannot be certified. Based on the same principles as the neural network recognition of hand written characters, our machine learning approach assumes no particular underlying theory or parameter values; it only assumes that the training sets are representative of the signals that will later be presented to the algorithm.

\begin{figure}
\centering
\includegraphics[width = 0.98\linewidth]{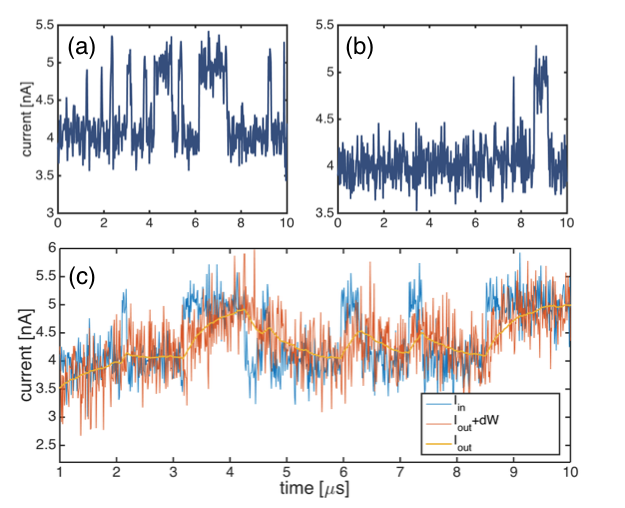}
\caption{Panels (a) and (b) show examples of Markovian QPC current with Rabi frequencies, $\Omega=\gamma_\uparrow=5$, and tunneling rates, $\gamma_\downarrow=4$ and $\Omega = 4$, $\gamma_{\uparrow,\downarrow}=1$. The panel (c) shows the application of the non-Markovian filters on the Markovian current of the type shown in (a) and (b).}
\label{fig:currents}
\end{figure}

\begin{figure*}
\centering
\includegraphics[width = 0.95\textwidth]{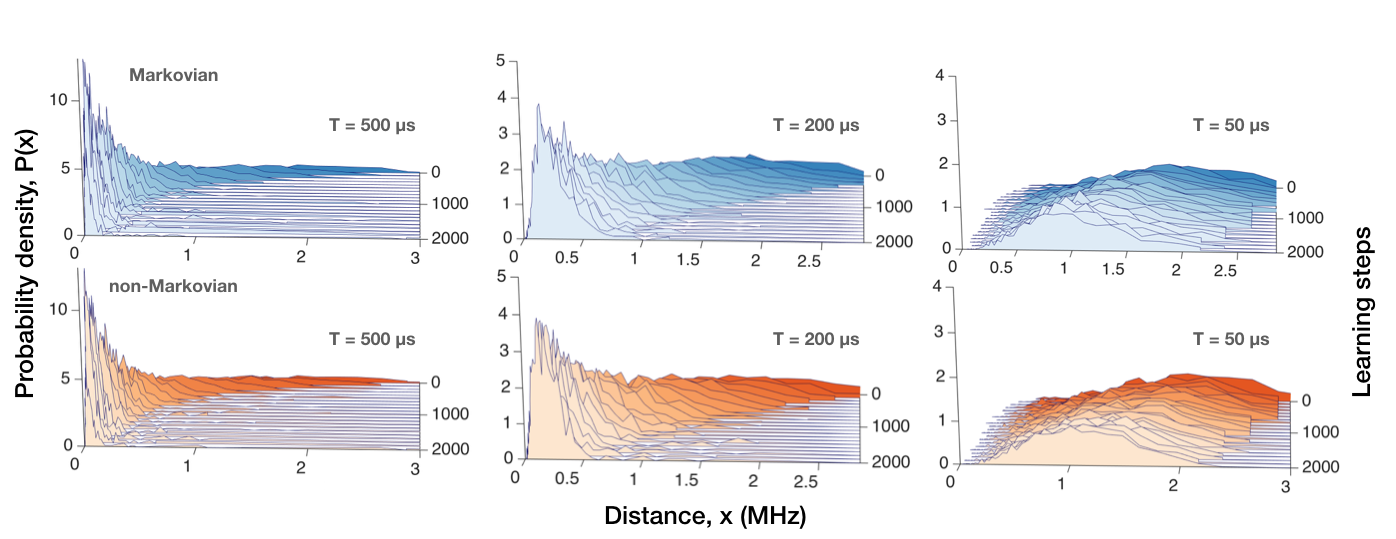}
\caption{Probability density of the distance from the correct value of all parameters given for different number of learning steps. The upper (lower) panel shows the first $2000$ learning step of learning based on simulated experimental currents from a markovian (non-markovian) data.}\label{fig:histograms-all}
\end{figure*}

To illustrate the flexibility of machine learning techniques for parameter estimation of quantum systems, we consider the continuous monitoring of a single electron that may tunnel on and off a quantum dot~\cite{Greplova2017, RevModPhys.79.1217}. The energy splitting between the spin-up and spin-down states can be tuned by a magnetic field, such that only a spin-down electron may tunnel into the dot. When an electron occupies the dot, further charging of the dot is prevented by the Coulomb blockade. Similarly, the electron may only tunnel off the quantum dot when it occupies the spin-up state. In addition to this incoherent tunneling on and off the dot, governed by the tunneling rates $\gamma_\downarrow$ and $\gamma_\uparrow$, respectively, the spin may precess between the spin states with a Rabi frequency $\Omega$ when an external resonant drive is applied. The electron is, thus, subject to both incoherent and coherent dynamics and is simultaneously monitored by a nearby QPC, which permits a noisy transmission current with a mean rate that depends on the charge on the quantum dot.  
The QPC current is insensitive to the spin state of the electron~\cite{Elzerman2004,Petta2180}. The left hand side of Fig. \ref{fig:NN} and the panels (a), (b) in Fig. \ref{fig:currents} show simulated QPC current data and clearly witness periods of low and high transmission, pertaining to the occupied and unoccupied states of the quantum dot. From the data shown, it becomes evident that one may apply a statistical analysis of the duration of low and high transmission intervals and, thus, obtain an estimate of the tunneling rates and the Rabi frequency. Assuming the validity of the stochastic master equation and the resulting Bayesian filter, an effective and precise estimation of the Rabi frequency and the tunneling rates from the QPC current was demonstrated in Ref. \cite{Greplova2017}, and is possible even when one cannot clearly discern low and high transmission intervals due to the signal noise.

\begin{figure}
\centering
\includegraphics[scale=0.55]{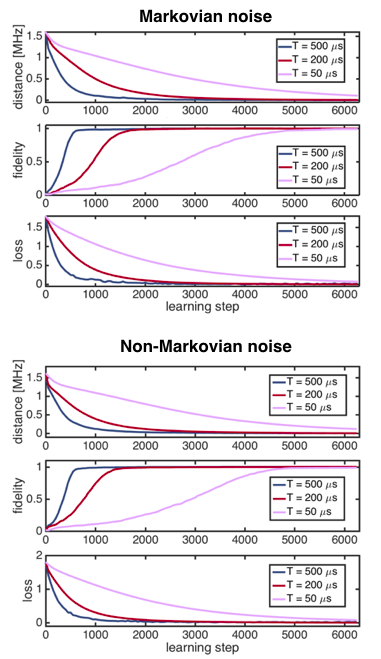}
\caption{Quantities characterizing learning success for three different lenghts of the time trace. We show average distance from the correct parameter, fidelity of our guesses and value of the loss function as a function of learning steps for Markovian (non-Markovian) dynamics in the upper (lower) panel.}\label{fig:plotfinal}
\end{figure}

To study the achievements of machine learning for the estimation of parameters leading to QPC signals as the ones shown in the figures, we synthesize many such signals, which we then treat as 1D image inputs for the convolutional neural network \cite{goodfellow2016, fei-fei-li} (CNN) as schematically shown in Fig. \ref{fig:NN}. 
To understand why image recognition is a suitable approach to parameter estimation consider a standard case for image recognition problems namely the recognition of hand-written images of digits between $0$ and $9$ \cite{lecun-mnisthandwrittendigit-2010, tensorflow2015-whitepaper}. A CNN, when trained, can assign any hand-written input to one of the $10$ basic classes. In our case, the classes are chosen as different combinations of parameters governing the dynamics of an experiment. Hand-writing from thousands of people include inherent randomness which the CNN learns to ignore and, similarly, the CNN learns to disregard irrelevant experimental noise fluctuations and extract only the features that distinguish the experimental parameters.

The neural network is shown in Fig. \ref{fig:NN} and consists of convolution layers that perform discrete convolution with a filter function, pooling layers that down-sample the data and dense fully connected layers that recognizes finer structure in the input data. The functionality and construction of these layers are described in detail in the Supplementary Material. The weights of the convolution filters and the fully connected layers define the connections between the neurons in the consecutive layers and are the parameters that are adjusted during the training of the CNN so as to minimize the so-called loss function (also called the cost function). For typical classification problems, the cost function is of the form
$H(y^{target},y^{output}) = -\sum_j y_j^{target}\ln y_j^{output}$,
where $y^{output}$ is the output of the network, and $y^{target}$ is the expected (correct) output. During the training, we re-evaluate the weights of the connection between the neurons in the network in order to minimize the distance between the output and the correct values~\cite{RDH1989,tensorflow2015-whitepaper}.

When the training is completed, we test the network to validate how well it performs on new signals. The learning and the validation may be done with purely experimental data, but in this work, both are numerically simulated. As an example, we consider $6$ candidate values for each of the parameters: the Rabi frequency $\Omega$ (equidistantly distributed between $4$~MHz and $10$~MHz) and the tunneling rates on and off the quantum dot $\gamma_{\uparrow,\downarrow}$ (equidistantly distributed between $1$~MHz and $6$~MHz). This gives us $6^3 = 216$ classes the experimental current can be assigned to. In the upper panel of Fig. \ref{fig:histograms-all}, the training of the network is visualized by the probability density of the average distance between the correct underlying parameters and the output parameters of the machine learning algorithm as a function of the number of learning steps. The numerical effort of the learning process depends on the total length, $T$, of the measurement record, and we considered signals of different duration, $50 \mu$s, $200\mu$s, $500 \mu$s, noting that the shorter intervals provide less information as witnessed by the broader range of estimation errors. Fig. \ref{fig:histograms-all} shows how the length of the trace influences the number of training sets needed for convergence towards the correct value of all parameters as a function of the number of learning steps (for the first $2000$ learning steps). When the network is trained, we perform validation on new signal currents (non-overlapping with the training data) and evaluate quantities that shows us the performance of the neural network as a function of the number of learning steps. In the three upper panels of Fig. 4 we show the average distance between the predicted parameters and the correct values, the fidelity (or accuracy) defined as the probability of predicting the right parameter and the loss function each as a function of the number of the learning steps (for details see Supplementary Material). Even though the convergence rate differs depending on the length of the time trace in all cases the CNN recognizes all parameters with  more than $99\%$ fidelity and the average distance from a correct value is on the order of $10^{-3}$~MHz for the longer time traces and $10^{-2}$~MHz for $50\,\mu$s (see Supplemental Material for precise values). While for the first $2000$ training steps the distribution of distances is quite broad (as shown in Fig. \ref{fig:histograms-all}), we see in Fig. \ref{fig:plotfinal}, that (albeit more slowly) the algorithm finds the correct value for all lengths of the time trace. Further technical and quantitative details of our CNN simulation are described in detail in the Supplementary Material.

So far, we considered a system amenable to a stochastic master equation analysis, but it is easy to extend our approach to realistic situations where a stochastic master equation does not apply. Assume for example that the QPC signal is modified by convolution with a filter function, $I_{out}(t) = (I_{in} \otimes u)(t) = \int_{-\infty}^{\infty} u(t') I_{in}(t - t') dt'$ where $I_{in\, (out)}$ is the ideal (actually detected) current. Due to the temporal convolution, we shall refer to the signal as non-Markovian. Moreover, we model the inclusion of further amplifier noise by adding an additional white noise $dW$ to $I_{out}(t)$ as shown in Fig.~\ref{fig:currents} (c). It is apparent that the step structure of the ideal current is now masked by the finite bandwidth of the filter together with the added random noise. Still, we may expect that the neural network may identify characteristic, albeit less intuitive, differences between the signals pertaining to the different parameter sets.

The CNN in Fig.~\ref{fig:NN} is now trained on the non-Markovian currents in the same way as described above, and it is interesting to note the close similarity of the peformance of the system in the Markovian and the non-Markovian case, cf. the upper and lower panels in Fig.~\ref{fig:histograms-all}. Also, the average distance, the fidelity and the loss functions, depicted in the lower panel of Fig. \ref{fig:plotfinal} for the non-Markovian currents follow the same evolution as for the Markovian case.  Despite a slight decrease of the speed of convergence in the non-Markovian cases, it is clear that qualitatively, the neural network treats the two problems with similar success. Thus, while no practical analytical theory exist to describe parameter estimation from the non-Markovian current signals, the machine learning approach allows us nevertheless to estimate the parameters with high accuracy.

In summary, in this letter, we proposed to use training of neural networks for the purpose of parameter estimation. We showed that this method is capable of dealing with uncertainties in, or complete absence of, models describing the experimental noise during measurements. In particular, our neural network approach translates a quantum parameter estimation problem into an image classification problem and therefore does not suffer from issues related to the characterization of quantum and classical noise. We imagine that our machine learning approach may have further applications, e.g., in connection with the use of the quantum systems for quantum communication and computing, where certain operations conditioned on measurement outcomes, feedback and error correction may be decided and executed based on the neutral network. Another daunting perspective  is the application of a quantum processor for the machine learning process itself which, apart from offering a potential speed-up, \cite{BiamonteQML,PhysRevLett.117.130501,ROA2017,SorenLearning}, may implement an almost autonomous quantum control and feedback systems with entangled system and observer.

\begin{acknowledgements}
EG and KM acknowledge financial support of Villum Fonden.
\end{acknowledgements}

\vspace{1 cm}
\begin{center}
\Large{Supplementary Material}
\end{center}

\appendix

\section{Introduction to Machine Learning}
\begin{figure}
\centering
\includegraphics[scale=0.33]{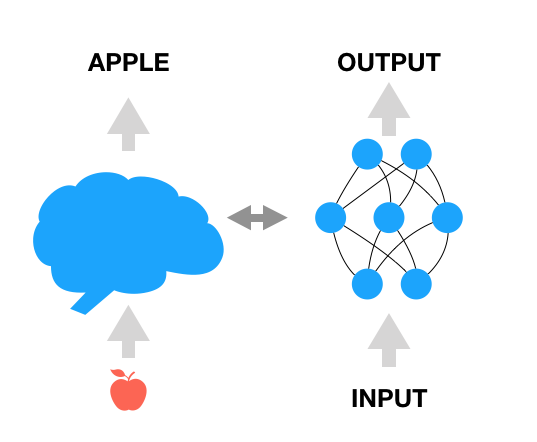}
\caption{Schematic of machine learning and neural network analogy.}\label{fig:brain}
\end{figure}

Machine learning generally stands for a wide number of techniques. In this section, based on \cite{florian} and \cite{Nielsen2015,goodfellow2016}, we give an introduction to one of the most successful machine learning techniques, artificial neural networks. The artificial neural networks are based on the idea of mimicking the functioning of the neuron connections of the human brain. As schematically shown in Fig. \ref{fig:brain}, this machine learning technique applies training in a fashion similar to human learning with the goal to be able to process complex inputs and conclude correct outputs.

\begin{figure}
\centering
\includegraphics[scale=0.3]{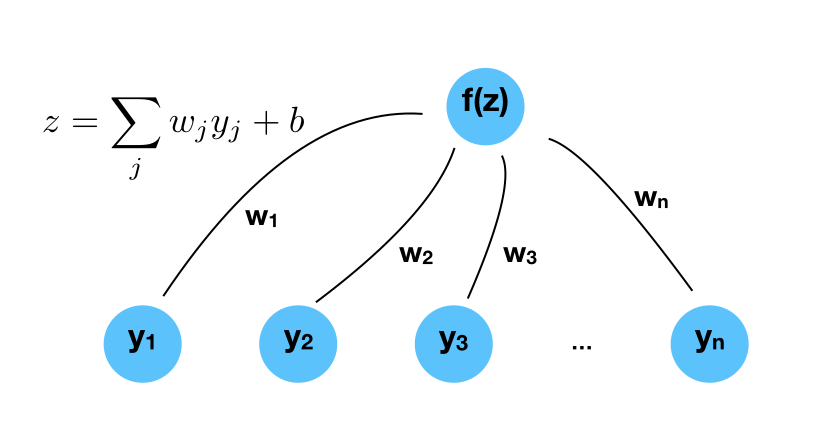}
\caption{Illustration of a single neuron within neural network.}\label{fig:OneNeuron}
\end{figure}

\begin{figure}
\centering
\includegraphics[scale=0.35]{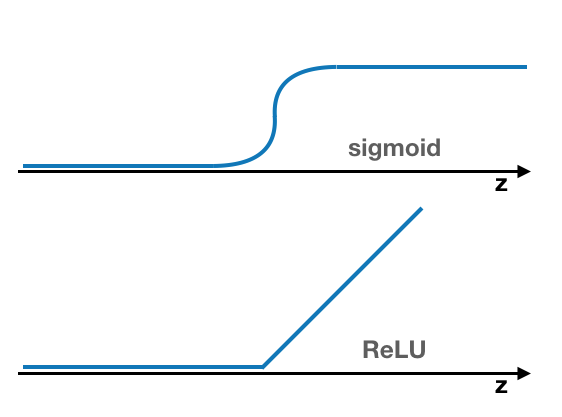}
\caption{Examples of the non-linear functions, $f$, applied to the result of linear operation, $z$.}\label{fig:nonlinear}
\end{figure}

A neural network is composed of neurons that are interconnected by strength parameters optimized by training of the neural network. 
Let us a consider a single neuron as shown in Fig. \ref{fig:OneNeuron}. This neuron is connected to $n$ neurons in the preceding layer $\{y_1,\dots,y_n\}$. The output of the neuron is given by a nonlinear function, $f$, applied on the weighted sum of inputs $y_i$,
\begin{equation}
\label{eq:neuron_linear}
z = \sum_j w_jy_j +b,
\end{equation}
where the weights $w_j$ represent the strength of a given connection and $b$ is a constant bias offset. Each neuron is equipped with its own bias and set of weights (these are the parameters that will be adjusted by training), while the nonlinear function is typically fixed for the whole network. Simple examples of the so-called activation function, $f$, are for instance the sigmoid function and the so-called rectified linear unit (ReLU) function shown schematically in Fig. \ref{fig:nonlinear}. The sigmoid function asymptotically approaches the Fermi function while the ReLU function is defined to return $0$ for negative inputs and input itself for positive ones.

The neural network can be trained by minimizing the cost function (also called the loss function). To quantify how `far' we are from the desired output we introduce cross-entropy as
\begin{equation}
\label{eq:cross-entropy}
H(y^{target},y^{output}) = -\sum_j y_j^{target}\ln y_j^{output}.
\end{equation}
Here, $y^{output}$ is the output of the neural network normalized as a probability and $y^{target}$ is a vector encoding the correct classification (the correct outputs are stored as one-hot vector, with $1$ at the position corresponding to the correct digit and $0$ elsewhere). The cross-entropy, $H$, is always non-negative and becomes zero only for $y^{target}=y^{output}$.

The minimization of the cross-entropy \eqref{eq:cross-entropy} for each training sample is done by means of gradient descent method \cite{ADAM}. To make this minimization computationally tractable, the so-called back-propagation algorithm is being used \cite{back-prop}. More precisely, the back-propagation allows us to approximate the derivative of the cost function by averaging over the training set. The details of the algorithm are explained in e.g. \cite{Nielsen2015}. A simple Python implementation of an example of this procedure using the Tensorflow library \cite{tensorflow2015-whitepaper} is shown in \cite{mnistT}.

When the training is finished, the validation set serves to establish the fidelity (the probability that given digit is recognized correctly).

\begin{figure}[b]
\centering
\includegraphics[scale=0.35]{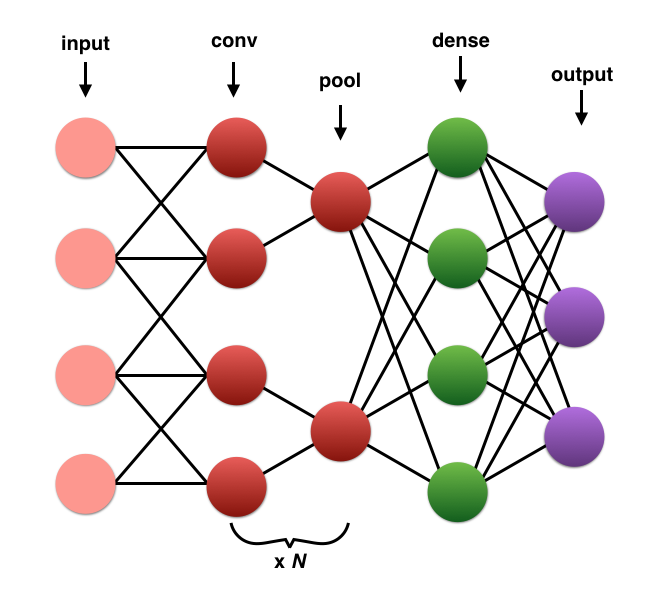}
\caption{Example of evolved neural network for image recognition: convolutional layer followed by pooling layer $N$ times connected with with dense layer.}\label{fig:all_layers}
\end{figure}

\begin{figure}
\centering
\includegraphics[scale=0.36]{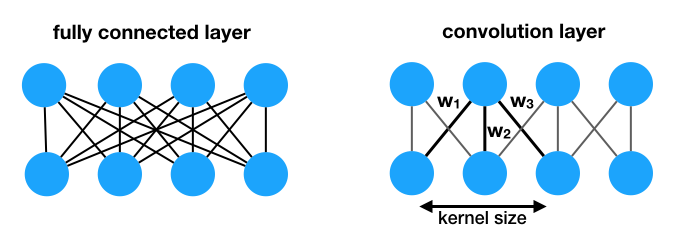}
\caption{Illustration of the difference betweem fully connected dense layer and convolution layer.}\label{fig:FvsC}
\end{figure}


\begin{figure}
\centering
\includegraphics[scale=0.35]{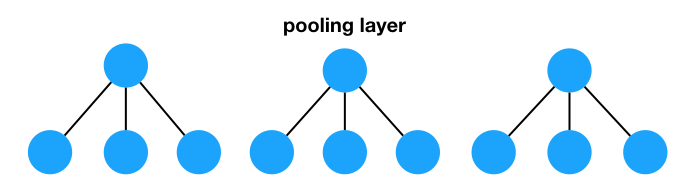}
\caption{Illustration of the functionality of the pooling layer.}\label{fig:pool}
\end{figure}

An important factor of the functionality of the neural networks is the way the layers of neurons are interconnected. An example of evolved artificial neural network is shown in Fig. \ref{fig:all_layers}. This particular structure of neural network is oftenused for image recognition problems.

A type of layer that is especially relevant for image recognition is the so-called \emph{convolution layer}, in which each neuron in the layer is only connected to the small neighborhood of neurons in the preceding layer (see Fig. \ref{fig:FvsC}). In practice multiple convolution layers are typically being used to address different features of input data. 
The technical aspects of convolutional neural networks are described in  \cite{fei-fei-li}. 
The layer used for down-sampling in the network is called the \emph{pooling layer}. The pictorial representation of pooling layer functionality is shown in Fig. \ref{fig:pool}.
In case all the neurons of the given layer have a connection to each neuron of the following layer, the layer is called \emph{fully connected dense layer}.



\section{Technical aspects of CNN training}

The neural network used in this work is a convolutional neural network consisting of an input layer, two convolution layers (followed by pooling), one densely connected layer and an output layer. The activation function for the convolution layers and the densely connected layer is of the ReLU type as explained in Appendix A. The input layer represent here the continuously measured signal and is thus discretized on a time axis into bins of length, $\Delta t$. We therefore have an input layer of $N = T/\Delta t$ nodes, where $T$ is the full time of the recorded measurement current. We set $\Delta t = 10\,\text{ns}$ which corresponds to a realistic sample rate for experimental realizations and the simulations are performed for up to $T=500\,\mu\text{s}$ such that we have $N=5 \times  10^4$. In other words, the input layer can be represented by a tensor with dimension $[M \times N \times 1]$, where $M$ is the \emph{batch size}, i.e. the number of currents loaded. The input data is scaled to be between 0~and~1.

The batch size is arbitrary as no optimization parameter depends on $M$. As such, once the network is trained, we can perform the prediction step with a single experimental current, i.e. $M=1$. For the training we used consecutive batches each with a batch size of $M=1000$ and for each batch we perform 250 learning steps (back-propagations) and similarly, in order to have a small statistical error, we evaluated the performance of the network with a single batch with $M=1000$. The evaluation batch was not used for training. The batches are generated offline, i.e.  prior to loading of the neural network, and to save memory usage only one batch is being saved in memory at a time. In the numerical simulation of the physical system, each trajectory used a different random seed and for all trajectories in the batch, the parameters were randomly chosen in order to sample the whole parameter space. For all the generated currents, we also save the parameters used in the simulation which we can use to generate the target probability function and for calculating the distance measured used in the main text. In practice, the target parameters are loaded into memory simultaneous with the currents and saved as a \emph{one-hot tensor} in each parameter that is a tensor representing the probability distribution for each parameter with only one non-zero entry.

The input layer is followed by a convolution layer with a kernel size of 9 such that each neuron is connected to 9 nodes in the input layer. We use 16 convolutional filters.
We follow the convolution with a pooling layer with a pool size of 2. The resulting tensor is therefore $[M \times N/2 \times 16]$ and the layer is represented by $9 \times 16$ parameters, which are updated and optimized in each learning step. The second convolution layer is similar but with 32 filter resulting in a tensor size of $[M \times N/4 \times 32]$.

After the two convolution layers, the network has a densely connected layer with 1024 neurons with a ReLU activation function. As a result, the tensor is of dimensions $[M \times 1024]$ and the layer has $1024 \times N/4 \times 32$ optimization parameters. During the training, the densely connected layer is followed by a so-called dropout layer such that, for each learning step, a fixed random fraction, 40\% in this work, of the neurons are dropped out of the network.

\begin{figure*}
\includegraphics[width = 0.85\linewidth]{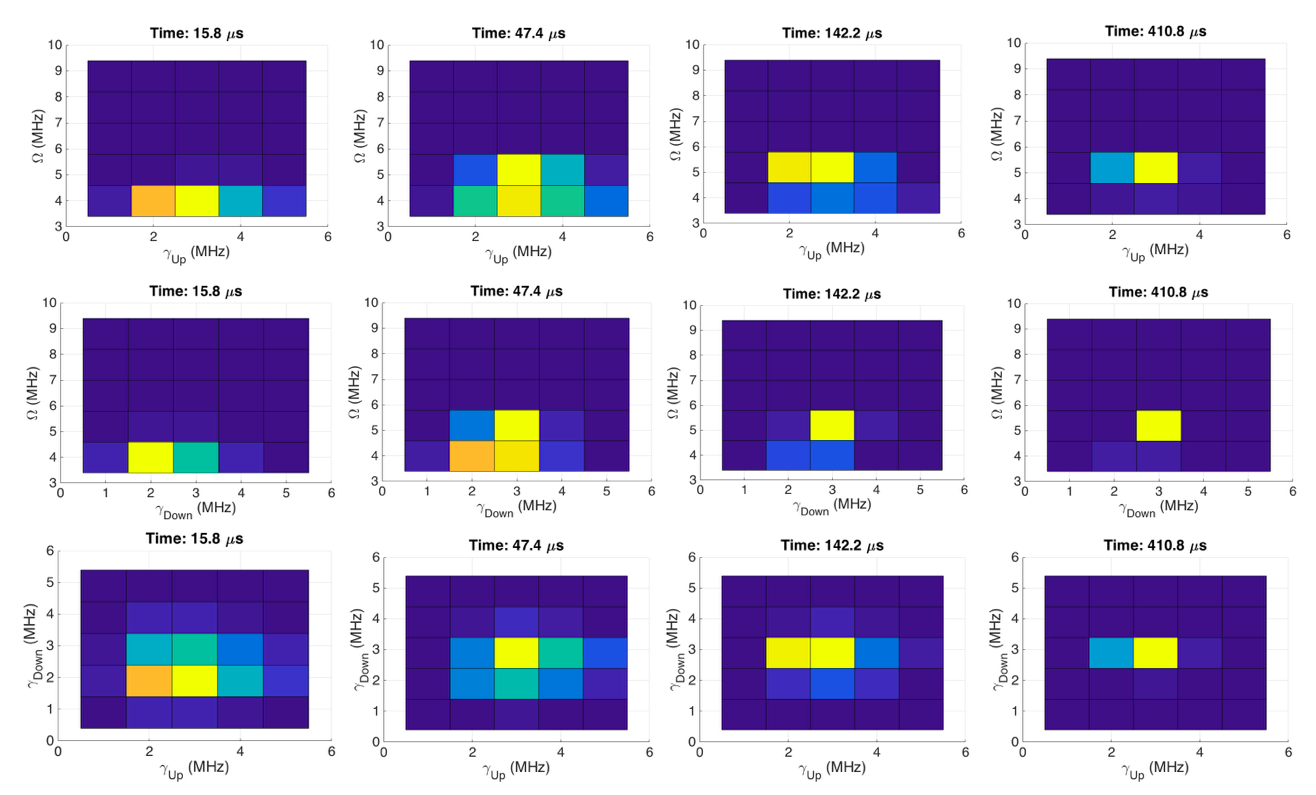}
\caption{Likelihood function projected onto two-dimensional subspaces shown for different lengths of the time trace. The correct parameters are $\lbrace \Omega,\, \gamma_\uparrow,\, \gamma_\downarrow \rbrace = \lbrace 5.2,\, 3,\, 3\rbrace$~MHz for this trajectory.}
\label{fig:bayes}
\end{figure*}

Since we have three parameters that we want to estimate each of which are picked from a set of 6 values, the output layer is of the size $[M \times 3 \times 6]$. A renormalization function is applied to each of the three dimensions such that the entry $[i,\,j,\,k]$ denotes the probability that the $i$th currents $j$th parameter has the $k$th value. The output layer is now finally compared to the target tensor. For the learning steps we calculate the cross-entropy for each parameter, while the evaluation steps output the average distance to the target and accuracy of the predictions.

The fidelity is calculated in a straightforward manner by considering the correct parameters for the trajectories in the evaluation batch. Here $p'_{i,j}(k) \in \lbrace{0,1\rbrace}$ denotes the probability vector for the $j$th parameter of the $i$th trajectory, but as the correct parameter is known $p'_{i,j}(k)$ is only non-zero for a single $k$. Similarly we denote the probability outcomes of the softmax activation $p_{i,j} (k)$. The function $\text{argmax}_k(p_{i,j})$ returns the $k$ corresponding to the maximal value of $p_{i,j} (k)$. We can therefore define the fidelity as
\begin{align}
F = \frac{1}{3 M} \sum_{i=1}^{M} \sum_{j=1}^3 \big[ \text{argmax}_k(p_{i,j}) == \text{argmax}_k(p'_{i,j}) \big],
\end{align}
with $x == y$ denoting the logical equal operation, which returns 1 if $x$ is equal to $y$ and otherwise 0.

\begin{table}[t]
\begin{tabular}{lrrr}
Average distance $\quad$ & $T=500\,\mu\text{s}$ & $T=200\,\mu\text{s}$ & $T=50\,\mu\text{s}$ \\
\hline
\hline
Markovian & 0.0028 MHz & 0.0064 MHz & 0.1084 MHz \\
Non-Markovian & 0.0033 MHz & 0.0050 MHz & 0.1166 MHz \\
\hline
\hline
\end{tabular}
\caption{Average distance after 6250 learning steps for the three total measurement times for both the Markovian and non-Markovian case.} \label{tab:distance}
\end{table}

The last figure of merit that we calculate to evaluate the performance of the parameter estimation is the average distance from the estimated value to the real value. The parameter values for the parameter estimation performed in this work are as follow: $\Omega \in \lbrace 4,\, 5.2,\, 6.4,\, 7.6,\,8.8,\,10\rbrace\, \text{MHz}$ and $\gamma_\uparrow, \gamma_\downarrow \in \lbrace 1,\, 2,\, 3,\, 4,\, 5,\, 6 \rbrace \, \text{MHz}$. Consider now the vectors $\vec{\Omega}$, $\vec{\gamma}_\uparrow$ and $\vec{\gamma}_\downarrow$ to contain the candidate values for each parameters. The average distance is now calculated as
\begin{align}
\bar{d} =&\, \frac{1}{3 M} \sum_{i=1}^{M} \Big( \big| p_{i,1} \cdot \vec{\Omega} - p'_{i,1} \cdot \vec{\Omega}\big| + \big| p_{i,2} \cdot \vec{\gamma}_\uparrow - p'_{i,2} \cdot \vec{\gamma}_\uparrow\big| \nonumber \\&+\big| p_{i,3} \cdot \vec{\gamma}_\downarrow - p'_{i,3} \cdot \vec{\gamma}_\downarrow\big| \Big),
\end{align}
with $x\cdot y$ denoting the inner product between $x$ and $y$.

The learning process was terminated at the point of convergence for the Markovian case with a time trace of $T = 500\, \mu\text{s}$ at which point the average distance was 0.0028 MHz. Similarly, the learning process for the rest of the simulated cases was terminated after the same amount of time-steps. The resulting average distance at termination is shown in Table.~\ref{tab:distance}.

\section{Comparison with Bayesian approach}
For the case where the Markovian approximation holds, the underlying parameters governing the quantum dynamics can be estimated using Bayesian methods. For the situation considered in the main text, this leads to the task of solving a stochastic master equation for 216 sets of parameters for each trajectory in order to estimate the most likely parameter. 

The result of the master equation simulations is a likelihood function $L_t(i,j,k)$ indicating the likelihood at time $t$ for the candidate values given by the indices's $i$, $j$ and $k$ corresponding to $\Omega$, $\gamma_\uparrow$ and $\gamma_\downarrow$ respectively. Normalizing the likelihood function at each time $t$ gives the probability distribution of the three variables, $P_t(\Omega, \gamma_{\uparrow}, \gamma_\downarrow)$. In Fig. \ref{fig:bayes} we show the marginal distributions of $P_t$ at 4 different times for a single trajectory. The most likely set of parameters converges as expected to the correct underlying set of parameters.

\bibliography{BibML.bib}
\end{document}